# Applications of Fractional Calculus to Diffusion Transport in Clay-Water System

**Dean Korošak[a], Bruno Cvikl[a,b], Janja Kramer[a], Renata Jecl[a], Anita Prapotnik[a]**

**and Miran Veselič[c]**

[a] Chair for Applied Physics, Faculty of Civil Engineering,
University of Maribor, Smetanova 17, SI-2000 Maribor, and

[b] "J. Stefan" Institute, Jamova 19, SI-1000 Ljubljana

[c] Agency for Radwaste Management, Parmova 53, SI-1000 Ljubljana

dean.korosak@uni-mb.si

## ABSTRACT

The analysis of the low-frequency conductivity spectra of the clay-water mixtures is presented. The conductivity spectra for samples at different water content values are shown to collapse to a single master curve when appropriately rescaled. The frequency dependence of the conductivity is shown to follow the power-law with the exponent $n$=0,67 before reaching the frequency-independent part. It is argued that the observed conductivity dispersion is a consequence of the anomalously diffusing ions in the clay-water system. The fractional Langevin equation is then used to describe the stochastic dynamics of the single ion.

## 1 INTRODUCTION

Numerous studies have already demonstrated and analyzed the anomalous properties of the diffusion transport in complex environments caused by the heterogenous nature of diffusivity [1]. This heterogenity stems from spatial distribution of different minerals, volume size, geometry, connectivity and distribution of pore space in the medium, and processes at the pore surfaces. Even when the diffusion coefficients are scaled to include for example tortuosity or sorption effects, the transport properties cannot be described with ordinary diffusion equation which is also true for diffusion in geometrically complex matter described as a fractal. Recently have several studies considered the fractal models to describe the geometric relations in soils and also properties of water in soils [2,3,4,5]

In relation to transport of contaminants through porous media (such as soil or rock) these anomalous features stem from motion of particles with the flow of water in saturated soils and from interaction with the porous solid matrix. In a performance assessment of a nuclear waste repository the release of radionuclides over long or very long times needs to be considered therefore the diffusion processes into solid matrix are important [6,7].

The main consequence of anomalous diffusion of radionuclides in geological formations surrounding the nuclear waste repository site is the time dependence of the diffusion coefficient. It has been suggested that the diffusion coefficient is a decreasing function of time [5], therefore on long time scale the assessment of diffusion processes will result in over prediction of diffusion into the surroundings and underestimate of the transport velocity.

The microscopic properties of diffusion transport of ions in also clayey soils [8] present important issues in migration studies of radionuclides in moist soil for soil decontamination and engineering of nuclear waste repository natural barriers. The effective diffusion constant was considered in [9] where the diffusion coefficients for $^{85}$Sr, $^{131}$I and HTO were obtained, while the dependence of the diffusion coefficient on the distance from the clay-water interface was studied in [10] showing the increase in diffusion coefficient near the interface with respect to its value in the bulk as much as ten times.

From the experimental part the traditional liquid phase through diffusion experiments are extremely time demanding. The alternative experimental techniques have therefore been proposed to characterize diffusion processes in porous media. In [11] the formation factor which characterizes the diffusive paths in porous medium was obtained utilizing electromigration processes [9,12,13] rather than diffusive and measuring the conductivity of the wet porous samples. However, the analysis of the results of electrical and dielectrical characterization of wet porous media turns out to be a very involved issue, specifically in the low frequency region due to the observed dispersion of the conductivity [14,15].

In this paper we present the analysis of the low frequency part of the conductivity spectra of the kaolinitic clay samples prepared at different water content approximately covering the range between plastic and liquid limits. We show that the measured spectra exhibit the anomalous part in the low frequency region displayed as a power-law frequency dependence before reaching the plateau. Furthermore, the conductivity dispersion curves for samples with different water content when rescaled collapse to a single master curve signalling the universality as frequently observed in disordered matter [16]. The attempt to explain the experimentally observed features is presented considering the stochastic motion of the ion near the clay particle-water interface.

## 2 EXPERIMENTAL

The electrical characteristics of the clay-water system were measured using the low frequency impedance analyzer at room temperature. The admittance of the sample placed in the measuring cell between two planparallel electrodes (area $S$=5,5 cm$^2$, distance $L$=4,5 – 5 mm) was determined from the linear response of the sample to the small oscillating bias on the electrodes of magnitude 10 mV. The real (conductance) and imaginary part (capacitance) of the admittance were measured in the frequency interval 100 Hz to 100 MHz.

The clay sample used was kaolinite intended for basic research purposes so the mineral composition and structural formula of the sample were known. The plastic and liquid limits of the sample were at 25,9 % and 40,1 % water content respectively. The specific surface of the clay sample was 10 m$^2$/g and its specific gravity 2,6.

The frequency dependence of the conductance and capacitance was determined first for dry sample, and then for wet samples, with water content ranging from 32 % to 56 %. Wet samples were obtained with addition of distilled water to the clay.

## 3 RESULTS AND DISUCSSION

The results of the measured conductivity as a function of frequency are presented in fig. 1. The water content of the measured samples was 32 % (the lowest conductivity dispersion curve in fig.1), 36 %, 40%, 48 %, 52 % and 56 % (the topmost dispersion curve in fig.1). All the samples show qualitatively similar behaviour. The conductivity increases for low enough frequencies and gradually obtains approximately constant value.

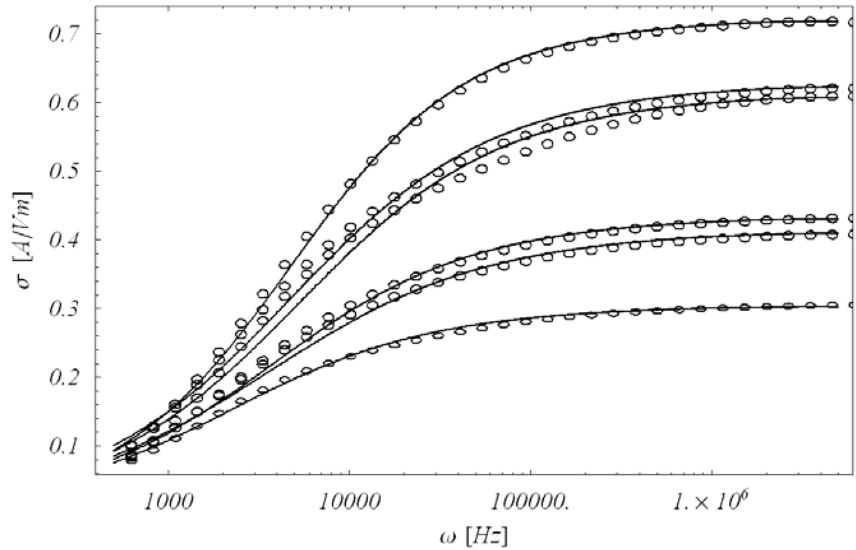

Figure 1: Measured (open dots) and calculated (solid lines) frequency dependence of conductivity of kaolinitic clay samples at the following values of water content: 32 %, 36 %, 40%, 48 %, 52 % and 56 %. The

The frequency dependence of the complex conductivity for the samples with different water content can be well described with the following expression:

$$\sigma_{lf}(\omega) = \frac{\sigma_0}{1 + (i\omega\tau)^{-n}}, \quad (1)$$

where $\sigma_0, \tau,$ and $n$ are constants determined from fitting of the real part of the expression given by eq. (1) for each curve. The conductivity given by eq. (1) follows the power-law $\sigma \propto \omega^n$ in the low-frequency region where $\omega\tau \ll 1$, and gradually approaches the constant part $\sigma_0$ for higher frequencies. The values obtained for the calculated curves shown in fig. 1 are given in table 1.

Table 1: Values of the parameter used to calculate the real part of the low-frequency conductivity shown in fig.1.

|  | 32 % | 36 % | 40 % | 48 % | 52 % | 56 % |
|---|---|---|---|---|---|---|
| $\sigma_0$ [A/Vm] | 0,31 | 0,41 | 0,43 | 0,61 | 0,63 | 0,72 |
| $\tau \times 10^4$ [s] | 2,5 | 2,8 | 2,7 | 1,9 | 2,1 | 2,1 |
| $n$ | 0,6 | 0,57 | 0,6 | 0,58 | 0,6 | 0,66 |

Using the limiting values $\sigma_0$ as a measure for the conductivity of the clay-water system at given water content, and plotting these values against the water content we obtained the graph as shown in figure 2.

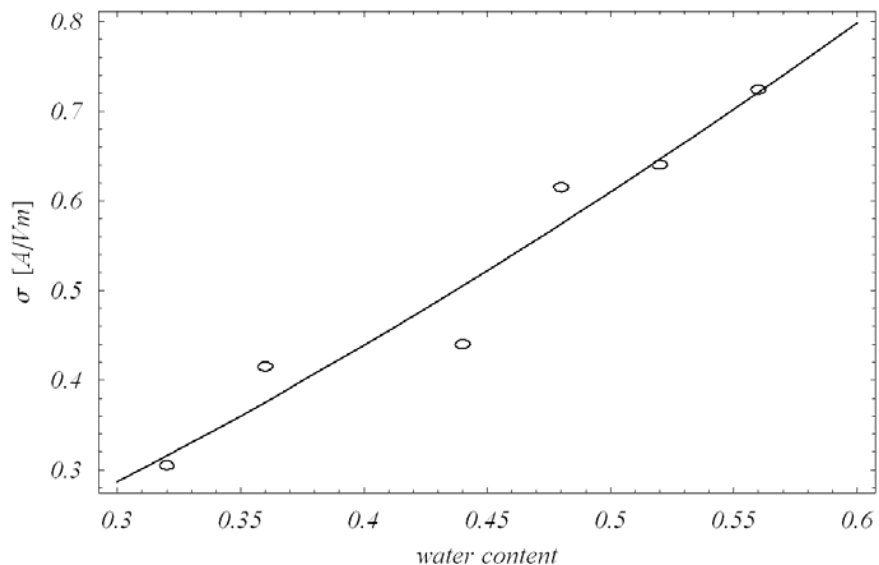

Figure 2: The dependence of the conductivity plateau on water content in clay-water system (open dots). The solid line is the fit with the function: $\sigma_0(\theta) = a\theta^m$, where $a$=1.69 A/Vm, and $m$=1.47.

The solid line in the figure 2 is the best fit to the conductivity vs. water content data using $\sigma_0(\theta) = a\theta^m$ as a fitting function. The value of the exponent fitted to the data in fig.2 was $b$=1,47 which is close to the value for moist brick (=1,6) obtained in the impedance spectroscopy analysis of the selected building materials [17] and for surface conduction in cement mortar (=1.3-1.5) [20]. Such dependence on the water content was attributed to the

004.5

emergence of the water isles in the pore space with the diminishing water content causing the lengthening of the diffusion paths of the ions.

The low frequency dispersion in the dielectric spectroscopy experiments is commonly assigned to electrode effects caused by the electrochemical processes similar to those occuring at the interface between the metal electrode and electrolyte [18,19] and dealt with invoking various equivalent circuits. The typical feature of the electrode polarization in the dielectric spectroscopy experiment is the large enhancement of the real part of the dielectric function with diminishing frequency and its dependence on electrode separation. The conductivity has in the same frequency interval showed to be frequency independent [19]. Berg et al. [20] have showed by detailed study of the dielectric properties of cement mortars at different water contents that the low-frequency part of the spectra consists of the electrode polarization part followed (in the frequency scale) by the dispersion attributed to the surface conduction in the pore space. They described the frequency dependence of the pore surface conductivity part with a power-law $\sigma(\omega) = C\theta^m \omega^n$, where $C$ is a constant. Exponent of the water content $m$ was in the range of $m = 1,3 - 1,5$ and the frequency exponent was found close to $n=0,7$ [20]. As showed above in figs. 1 and 2, and in table 1, our own findings on kaolinitic clay correlate surprisingly well with the results of Berg et al..

The power-law behaviour of the low frequency limit of eq. (1) indicated that the conductivity spectra at different levels of water content should scale to a single curve. The scaling of the conductivity of our samples is confirmed by fig. 3 which displays the log-log plot of scaled conductivity data for the samples with the lowest (32 %), intermediate (40 %), and highest (56 %) level of water content.

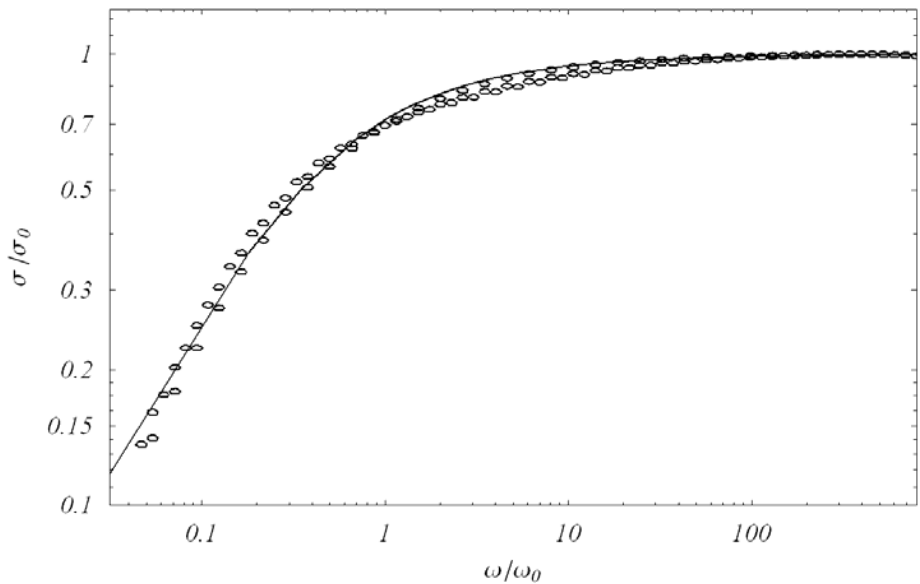

Figure 3: The log-log plot of the scaled frequency dependent conductivity obtained from the data for the samples with 32 %, 40 %, and 56 % water content, and the calculated conductivity according to the eq. 1 with $n = 0,67$.

The frequency dependence of the conductivity of the whole range of water content level in the clay-water mixtures can be described by a single master curve:

$$\sigma/\sigma_0 = 1/(1+(i\omega/\omega_0)^{-n}), \tag{2}$$

where the scaling frequency $\omega_0$ for each set of data was determined from the equation $\sigma(\omega_0) = \sigma_0/2$ as suggested by eq. (2). The value of the exponent $n$ for the data shown in fig.3 was found to be $n=0.67$.

Next we turn to the dynamical origin of the low frequency power-law behaviour of the surface conductivity $\sigma \propto \omega^n$ as given by the eq. (2). The properties of the diffusive motion of the ions near the clay particle surface are contained in the frequency dependent diffusion constant $D(\omega)$ to which the conductivity is connected via generalized Einstein relation:

$$\sigma(\omega) = \frac{q^2 \rho}{kT} D(\omega), \tag{3}$$

where $\rho$ is the ion density, $T$ temperature, $q$ elementary charge, and $k$ is the Boltzmann constant. The frequency dependent diffusion coefficient is given by [21]:

$$D(\omega) = -\frac{1}{6}\omega^2 \int_0^\infty e^{-i\omega t} \langle x^2 \rangle, \tag{4}$$

where $\langle x^2 \rangle$ is the mean-square displacement of the ion. Normal diffusion is characterized by the mean-square displacement that is linear in time $\langle x^2 \rangle \propto t$, while anomalously diffusing particles display mean-square displacement of the form $\langle x^2 \rangle \propto t^\beta$, where $\beta \neq 1$. The low frequency limit of the ion surface conductivity in clay-water mixture is therefore the consequence of the diffusive motion of the ions with $\langle x^2 \rangle \propto t^{1-n}$. It is however a non-trivial task to determine the correct underlying diffusion process since different dynamic processes yield the same mean square time dependence as for instance do fractional Brownian motion and fractal time process [22]. In case of diffusion in fractal space, as it is the case in porous media, it seems that the dynamical processes are described by the fractional Langevin equation [23,24]:

$$_0D_t^p v(t) = -\gamma v(t) + F(t) \tag{5}$$

Here $\gamma$ describes the viscosity of the pore solution, $F(t)$ is the stochastic force, and the time fractional derivative of the velocity $v$ of the particle is given with the Riemann-Liouville operator of the order $p$ [25]:

$$_0D_t^p f(t) = \frac{1}{\Gamma(1-p)} \frac{\partial}{\partial t} \int_0^t \frac{f(\tau)}{(t-\tau)^p} d\tau, \tag{6}$$

The asymptotic solution for long times of the mean square displacement is linear if the random force is the usual Gaussian δ-correlated process with zero mean. However, if the random force is given as a fractional derivative of the δ-correlated process describing the fractal environment:

$$F(t) = {}_0D_t^{1-p} g(t), \qquad (8)$$

the asymptotic behaviour of the mean square displacement is shown to be [24]:

$$\langle x^2 \rangle \propto t^{2p-1} \qquad (9)$$

Comparing this result with the previously derived directly from the analysis of the conductivity spectra we can connect the power-law exponent of the conductivity dispersion to the order of the fractional derivative in the fractional Langevin equation: $n = 2(1-p)$. In the case of clay-water mixture analyzed here we have $n=p=2/3$ so the ion surface diffusion transport is subdiffusive, and could be characterized with the time-dependent diffusion coefficient $D(t) \propto t^{-n}$.

## 4   CONCLUSIONS

The low frequency part of the conductivity frequency dependence of the clay-water mixtures at different water content levels are analyzed. It is shown that the origin for the anomalous conductivity dispersion can be attributed to the surface conductivity in the pore space. The scaling of the conductivity spectra is established, and the master curve is shown to follow the power-law in the low frequency part with the exponent equal to $n=0.67$. The dynamical origin of the surface conductivity of the pore space is described with the stochastic motion of the ion governed by the fractional Langevin equation, the order of the fractional derivative being equal to 1-$n$/2. The transport of the ions near the clay particle surface is found to be subdiffusive with the effective time-dependent diffusion constant $D(t) \propto t^{-n}$.

We have shown that the analysis of the low-frequency part of the dielectric spectra can yield important insight into certain properties of the diffusion processes of ions in porous media, and in relation to contaminant transport contribute to better understanding and prediction of the long time migration of radionuclides from nuclear waste repository.

**ACKNOWLEDGMENTS**

The authors wish to acknowledge the support of Slovenian Research Agency (Grant L2-6672-0797).